\documentclass{article}
\usepackage{spconf,amsmath,graphicx}
\usepackage{booktabs}
\usepackage{xcolor}
\usepackage{color}
\usepackage{etoolbox}
\usepackage{tikz}
\usepackage{float}
\usepackage[stable]{footmisc}
\usepackage{url}
\usepackage{subcaption}
\usepackage{nth}
\usepackage[inline]{enumitem}
\usepackage{algorithm2e}
\usepackage{romannum}
\usepackage{numprint}
\npthousandsep{,}

\title{Towards measuring fairness in Speech Recognition: Casual Conversations Dataset Transcriptions}
\name{%
\begin{tabular}{@{}c@{}c@{}c@{}c@{}}
Chunxi Liu$^{\dagger}$ \qquad Michael Picheny$^{\dagger}$\thanks{The authors would like to thank Zhe Liu and Weiyi Zheng  at Facebook for their assistance with this work.}  \qquad Leda Sar{\i}$^{\dagger}$ \qquad Pooja Chitkara$^{\dagger}$ \qquad Alex Xiao$^{\dagger}$  \\
Xiaohui Zhang$^{\dagger}$ \qquad Mark Chou$^{\dagger}$ \qquad Andres Alvarado$^{\dagger}$ \qquad Caner Hazirbas$^{\dagger}$ \qquad Yatharth Saraf$^{\dagger}$ 
\end{tabular}
} 
\address{$^{\dagger}$Facebook AI, USA}

\begin{document}
\ninept
\maketitle
\begin{abstract}
It is well known that many machine learning systems demonstrate bias towards specific groups of individuals. This problem has been studied extensively in the Facial Recognition area, but much less so in Automatic Speech Recognition (ASR). This paper presents initial Speech Recognition results on ``Casual Conversations" – a publicly released 846 hour corpus designed to help researchers evaluate their computer vision and audio models for accuracy across a diverse set of metadata, including age, gender, and skin tone. The entire corpus has been manually transcribed, allowing for detailed ASR evaluations across these metadata. Multiple ASR models are evaluated, including models trained on LibriSpeech, \numprint{14000} hour transcribed, and over 2 million hour untranscribed social media videos. Significant differences in word error rate across gender and skin tone are observed at times for all models. We are releasing human transcripts from the Casual Conversations dataset to encourage the community 
to develop a variety of techniques to reduce these statistical biases. 
\end{abstract}
\vspace{-0.05cm}
\vspace{-0.2cm}
%
\section{Introduction}
\label{sec:intro}
\vspace{-0.05cm}
The problem of algorithmic bias in machine learning (ML) systems is generally well known and well studied. When performance on certain groups of individuals are specifically impacted, a perception of unfairness can result. In the context of ML, the term ``Fairness" often refers to various attempts at correction of such errors or statistical biases. There is particular emphasis on advancing Fairness with respect to variables correspond to societally sensitive characteristics or traditionally marginalized communities, such as gender, ethnicity, sexual orientation, disability, etc.

Some particularly publicized examples of bias in machine learning systems have occurred in the area of Facial Recognition,
when it was realized that performance of multiple commercially available  systems were demonstrated to be much poorer on individuals with certain skin tones and also as a function of gender \cite{buolamwini2018gender}. 
It therefore seems reasonable to assume that if Facial Recognition systems are prone to unfairly distributed errors, then other systems that involve human-machine interactions might also be prone to similar shortcomings.
In particular, in the context of Automatic Speech Recognition (ASR), several studies have analyzed gender, race and dialect bias. in articles appearing in various media outlets \cite{hbr,forbes,businessinsider,soundhound,uxmatters,racistalgos}. Summarizing the findings, there are ASR performance differences across gender, race, and dialect which amplify the historical biases previously known to exist. This creates an impetus for the ML community to devise remedies, e.g. \cite{hbr} explicitly states that ``Everyone deserves their voice to be heard''.   

The focus of this paper is to announce the release of publicly available data that can be specifically used to investigate some of these issues in the context of ASR, and present preliminary results investigating such issues. Specifically, we are augmenting the already released ``Casual Conversations" dataset \cite{Hazirbas_2021_CVPR} with accurate manual transcriptions that can be used to evaluate existing models with respect to various potential biasing factors and also train new models directly from the dataset itself with respect to these factors. 

``Casual Conversations" is composed of over 45,000 videos of approximately 1 minute duration each collected from a set of 3,011 participants, comprising approximately 846 hours of data. The videos feature paid actors who agreed to participate in the project. Participants casually speak about various topics and sometimes depict a range of facial expressions. They explicitly provided age and gender labels themselves.  Also, a group of trained annotators labeled the participants’ apparent skin tone using the Fitzpatrick scale. The videos were recorded in the U.S. with a diverse set of adults in various age, gender and apparent skin tone groups. These videos were originally intended to be used for assessing the performance of already trained models in computer vision and audio applications for the purposes permitted in the data user agreement. The agreement prevents a user from developing models that predicts the values of these labels, but one may measure performance of an algorithm as a function of these labels. 

The original data release of Casual Conversations did not contain any transcriptions, and thus could not be easily used to examine Fairness issues in the context of speech recognition. As such, we have had the data manually transcribed to permit such an evaluation to take place, and also provide the transcriptions to the broader speech community. 
We then performed a preliminary evaluation of speech recognition performance across different speech recognition models as a function of gender, age and skin tone labels. While the phenotypical property of skin tone may not be as operative a variable in measuring comparative error for ASR as it is for computer vision, we opted to leverage the existing labels from the Casual Conversations dataset due to the likelihood skin tone may correlate with other characteristics that could be drivers of disparate ASR performance.
The rest of the paper describes the transcription process followed, the models employed, presents recognition results, and draws some preliminary conclusions about bias and fairness in the context of the Casual Conversations data. 
\vspace{-0.25cm}

\section{Previous work in Fairness in Speech Recognition}
\label{sec:prev}
\vspace{-0.05cm}
As machine learning systems have been utilized more frequently in decision making, biases or unfairness of the outcomes of these systems have become an active research area recently \cite{dwork2012fairness,chouldechova2017fair,chouldechova2017fairer}. ASR which is one of the application areas of machine learning is also subject to these fairness concerns. For instance, it has been shown that there is a performance gap between male and female speakers \cite{tatman2017gender} as well as black and white speakers \cite{koenecke2020racial}. The reason for unfairness in many of these cases is attributed to representation in the training data, i.e. having limited amounts of training data for certain groups of subjects, e.g. less female speakers than male speakers in a speech corpus~\cite{garnerin2019gender}. A recent study \cite{feng2021quantifying} also confirmed the hypothesis that ASR systems can perpetuate the societal biases. In \cite{garnerin2021investigating}, a commonly used benchmark dataset, namely LibriSpeech \cite{panayotov2015librispeech}, is subsampled to investigate the impact of the amount of training data on the ASR performance. The conclusion of \cite{garnerin2021investigating} is that individual variability, irrespective of gender, affects the final performance more significantly, hence intra-variability within gender groups is also found to be an important factor to investigate. These studies mainly focused on analyzing the existing biases in ASR outputs. One approach to reduce ASR performance gaps between different sensitive groups is provided in \cite{sari2021counterfactually} which builds on the counterfactual fairness idea proposed in \cite{kusner2017counterfactual}.

Since most ASR datasets do not come with speaker attribute labels such as age, gender, dialect, etc, most of the previous studies evaluated fairness of the ASR systems on a limited number of corpora. There are recent efforts on curating data for under-represented groups such as African-Americans in speech studies \cite{kendall2020corpus} or for various demographic groups \cite{meyer2020artie}. 
The ``Casual Conversations'' dataset provides data from a larger set of subgroups than \cite{kendall2020corpus}, and also contains conversational speech rather than  read speech as in \cite{meyer2020artie}.

\section{Data Transcription}
\label{sec:data}
In order to produce transcriptions usable by the community, verbatim transcriptions were produced with various mark-ups. Specifically, hesitations and disfluencies (``uh", ``um"...) and repeated words were kept as spoken. Colloquial wordings (``gonna", ``sorta") were maintained. Non-speech sounds, like music and laughter, were tagged. Numbers were spelled out as words as spoken, as were emails and URLs. Common named entities made up of acronyms (e.g., NASA, USA) were left spelled as colloquially written, though. 

In terms of metadata related to the transcription, the text was punctuated. Long pauses were all indicated with the tag $<$no-speech$>$. Speaker turns associated with primary speakers (the interviewee) and secondary speakers (the interviewer, and occasional third parties who were not the official interviewee) were marked and time-stamped. 
An example of a transcript produced following the above process is:

\noindent { \it [0.000] [secondary\_0.240\_secondary] would you rather work from home, or in an office and why? [/secondary\_2.903\_secondary/] $<$no-speech$>$   [3.890] 
 [primary\_4.183\_primary] um  $<$no-speech$>$  [7.345] I prefer a mix of both, because $<$no-speech$>$   [11.170] I like to have the structure of the office, $<$no-speech$>$  just to $<$colloquial$>$kinda$<$/colloquial$>$  create a routine, but I do prefer some days [18.010] being able to work from home, because it's just a  $<$no-speech$>$ [21.010] more convenient option, sometimes, when life gets busy. [/primary\_23.512\_primary/] 
  [secondary\_23.655\_secondary] mhm.  $<$no-speech$>$  $<$spk\_noise$>$ alright. [/secondary\_25.515\_secondary/] [25.560] 
}

In total, 846.1 hours of manual transcriptions have been produced.
Many of the recordings contain both speech from the primary speaker and the interviewer, and sometimes a third speaker as well. Since the metadata annotations refer to each primary speaker, we first remove videos with no primary speaker speaking. Second, we convert and segment each video into audio files via manual time stamps, such that each resulting segment only contains the primary speaker’s speech. This leaves in total 572.6 hours of audio from the original 846.1 hours. After segmentation, the longest utterance is about 224 seconds.


\setlength{\tabcolsep}{0.129cm}
\begin{table*}[h!]
\caption{\label{tab:result_all}{\it WER results on the complete Casual Conversations dataset. Rel. gap either refers to the relative WER difference between female and male, or refers to the largest relative WER difference between all pairwise skin types, with the corresponding pair indicated in bold.
}
}
\centerline{ 
\begin{tabular}{  c |  c  |  c c c | c  | c c c  | c c c c  c c  | c }
\hline \hline
    &  overall  &  \multicolumn{4}{c|}{gender}           &  \multicolumn{3}{c  | }{age} &  \multicolumn{7}{ c}{skin type}    \\ 
model & WER  & female & male & other &  rel. gap & 18-30 &  31-45 & 46-85 & \Romannum{1} & \Romannum{2} & \Romannum{3}   & \Romannum{4}  &  \Romannum{5}  &  \Romannum{6} &  rel. gap  \\ 
\hline \hline
LibriSpeech & 34.3 & \textbf{31.8} & \textbf{37.1} & 60.0 & \textbf{16\%} & 36.1  & 35.1 & 30.9 & \textbf{27.5} & 30.9 & 34.6 & 34.0 & \textbf{37.5} & 37.2 & \textbf{37\%} \\
Video, supervised  & 13.9  & \textbf{11.9} & \textbf{16.3}  & 31.8 & \textbf{37\%}  & 13.9 & 13.9 & 13.8 & \textbf{11.2}  & 13.1  & 14.0  & 13.7 &  \textbf{15.0} & 14.8 &  \textbf{34\%} \\ 
Video, semi-supervised & 9.8 & \textbf{8.5} & \textbf{11.3}  & 24.0 & \textbf{32\%}  & 9.7 & 9.8  & 9.8  & \textbf{7.8} & 9.3 & \textbf{10.4} & 9.9  & 10.1 & 10.0 & \textbf{33\%} \\ 
Video, teacher & 8.6 & \textbf{7.5} &  \textbf{9.9}  & 21.6 & \textbf{33\%}  & 8.4 &  8.6 & 8.6 &  \textbf{6.9} & 8.3  & \textbf{9.0}  & 8.4   & 8.9 & 8.8  &  \textbf{30\%} \\ 
\hline
\# of hours  &  573   &   312 &  249  &  0.2  &   -   &  198   & 188  &   174  &   22  &   160  &  135  &   49   &  89    &  118    &  --  \\
\hline \hline
\end{tabular}}
\end{table*}
\begin{figure*}[h!]
    \centering
        \subfloat[LibriSpeech model.]{
        \includegraphics[width=5.9cm]{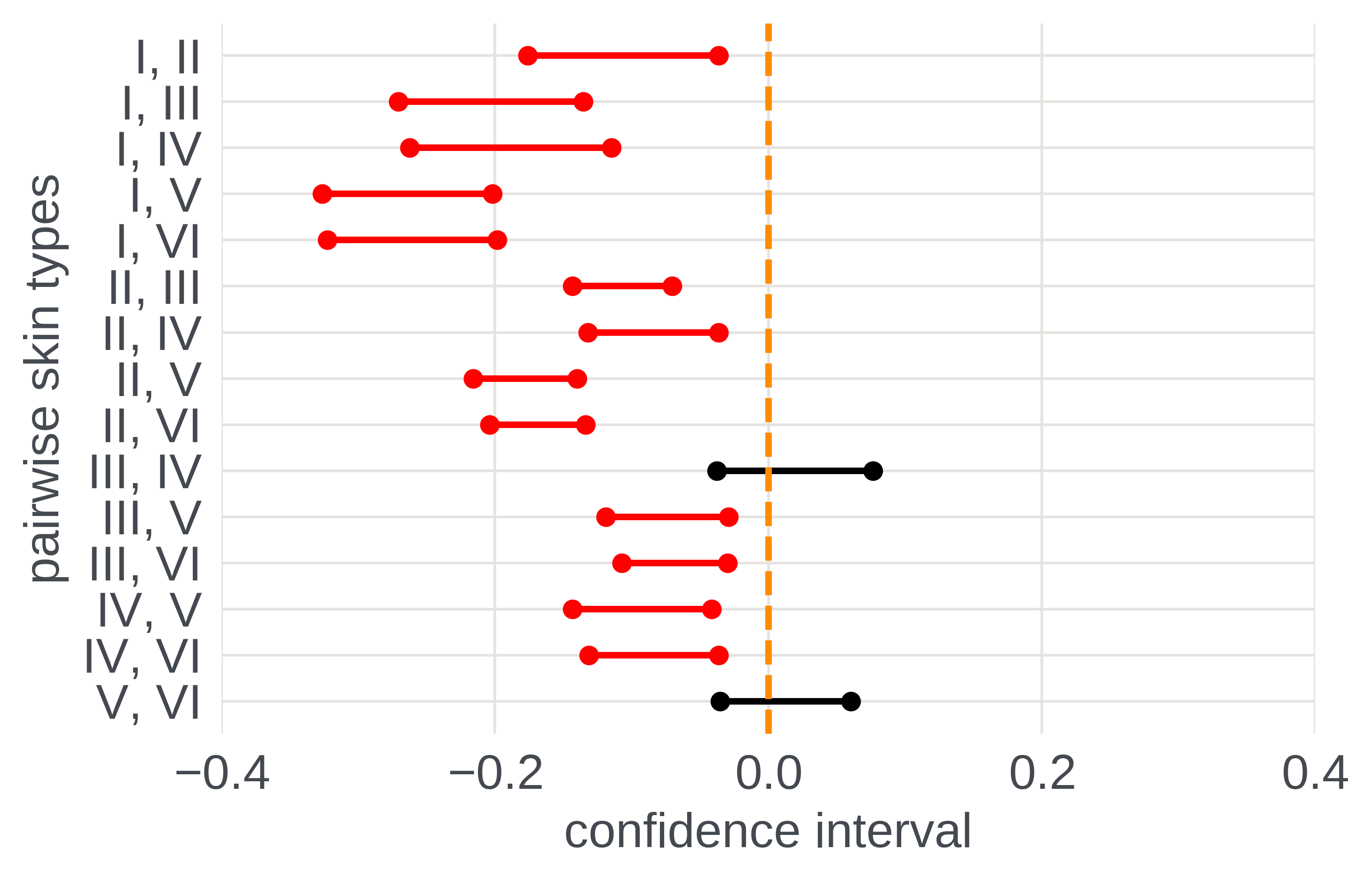}
        \label{fig:libri}    
    } \qquad 
    \subfloat[Video model, semi-supervised.]{
        \includegraphics[width=5.9cm]{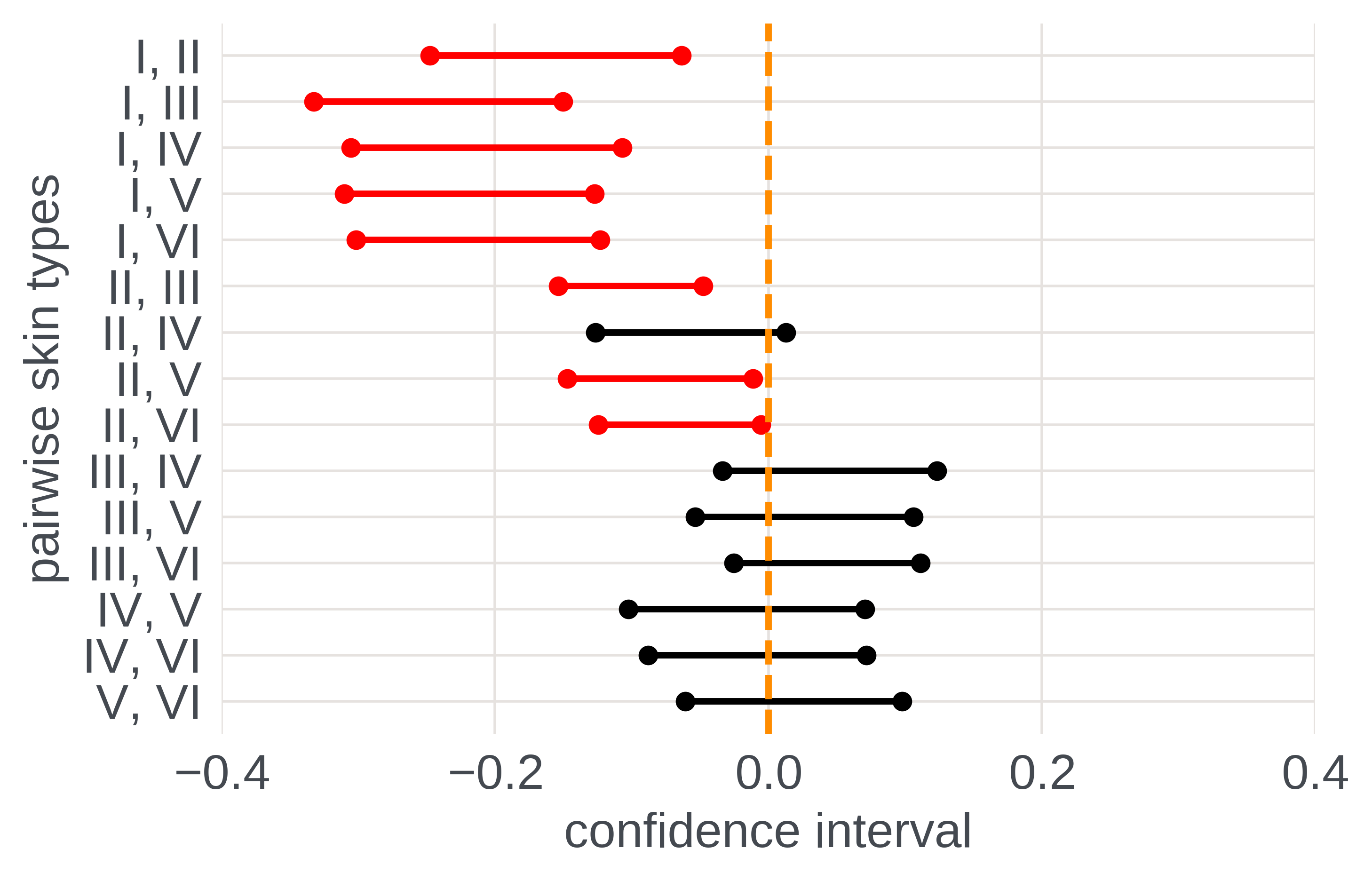}
        \label{fig:video_emformer} 
    }  
    \caption{Confidence intervals of each test statistic $\theta_{i, j}$  (Eq. \ref{eq:ratio}) for pairwise skin types. Each red line indicates the WER difference between two subgroups  statistically significant, and black     line insignificant.}
\label{fig:ci}
\vspace{-0.2cm}
\end{figure*}

\section{Speech Recognition Models}
\label{sec:model}

We built a series of recurrent neural network transducer (RNN-T) \cite{graves2012sequence} ASR models with respective sets of training data and configurations:
\begin{enumerate} [label=(\roman*)]    
\item  LibriSpeech model: a full-context conformer transducer model \cite{gulati2020conformer, yeh2021streaming} trained on LibriSpeech.
RNN-T output labels consist of a blank label and 1023 wordpieces generated by the unigram language model algorithm from SentencePiece toolkit \cite{kudo2018sentencepiece}.
Four 80-dimensional log Mel filter bank features are concatenated with stride 4 to form a 320 dimensional vector, followed by a linear layer and mapped to a 512 dimensional input to encoder.
Encoder has 17 conformer layers of embedding dimension 512, attention heads 8, feed-forward network (FFN) size 2048, and convolution kernel size 15.
Following \cite{li2021better}, we remove the original relative positional encoding, and reuse the existing convolution module for positional
encoding by swapping the order of convolution and multi-head self-attention modules.
Prediction network is a 2-layer LSTM of 512 hidden units and dropout $0.3$. 
Joint network has 1024 hidden units, and a softmax layer of 1024 units for blank and wordpieces. 
The word error rate (WER)  on \emph{test-clean}  and \emph{test-other} are $2.4$ and $5.5$ respectively. 
\item Video model, supervised: a streaming emformer model \cite{shi2021emformer} trained on 14K-hour manually transcribed social media videos. The video dataset is a collection of public and de-identified English videos, and contain a diverse range of speakers, accents, topics, and acoustic conditions.
The input feature stride is 6. 
Encoder network has 24 simplified emformer layers without memory bank, and each has embedding dimension 512, attention heads 8, and two macaron-like feed-forward network (FFN) modules \cite{gulati2020conformer} with FFN size 2048.   
Prediction and joint networks are the same as above, and 4095 wordpieces used instead. Model size is of about 140M parameters. 
\item Video model, semi-supervised: a semi-supervised streaming emformer model trained on over 2 million hour social media videos. 14K hours are manually transcribed as above, and the rest are unlabeled data and decoded by progressively larger teacher models.   
\item Video model, semi-supervised teacher: a final teacher model of one billion parameters and trained on over 2 million hour social media videos. 
\end{enumerate}
For all ASR training, we applied SpecAugment \cite{park2019specaugment}, alignment restricted training \cite{mahadeokar2021alignment} to improve training throughput, and auxiliary chenone prediction criteria to improve model convergence and performance \cite{liu2021improving}.
For all neural network implementations, we used an in-house extension of the PyTorch-based \emph{fairseq} \cite{ott2019fairseq} toolkit. 
All experiments used the Adam optimizer \cite{kingma2014adam}, tri-stage \cite{park2019specaugment} learning rate schedule with peak learning rate $4e^{-4}$, multi-GPU and the mixed precision training supported in \emph{fairseq}.

\section{Bootstrap Confidence Interval Method}
\label{sec:bootstrap}

In most speech recognition systems, the test data remains fixed and the nature of the speech recognition model used for decoding is the critical variable. Typical tests for statistical significance utilized for such comparisons are described in \cite{115546,266481}. However, in our case, the test data is different for each primary speaker. This implies a different test of statistical significance is needed. We decided to conduct significance tests using the bootstrap method \cite{efron1994introduction}, designed to compare data drawn from disparate populations that does not rely on the underlying assumption that the underlying data have a normal distribution.

Assume we would like to conduct statistical testing to determine whether a significant difference exists between the WERs of two subgroups  $i$ and $j$, where $i$ and $j$ can denote female and male, or any two skin types,  $\forall i, j =  \text{\Romannum{1}} \ldots \text{\Romannum{6}} $.  
WERs for subgroup $i$ and $j$ are denoted as $\text{WER}_{i}$, $\text{WER}_{j}$. 
Then we define the test statistic $\theta_{i,j}$ as the empirical WER ratio minus one:
\begin{equation}  
\vspace{-0.08cm}
  \theta_{i,j} = \frac{ \text{WER}_{i}  }{\text{WER}_{j}  }  - 1    
  \label{eq:ratio}
\vspace{-0.08cm}
\end{equation}
\noindent 
In the bootstrap method \cite{efron1994introduction}, each bootstrap sample is generated by the following process:
\begin{enumerate}   
\item  subjects are repeatedly sampled with replacement from the original subgroups $i$ and $j$, and the number of samples are equal to the number of subjects in respective original subgroup. Then $\text{WER}_{i}$ and $\text{WER}_{j}$ are calculated on the samples. 
\item a parameter estimate for $\theta_{i,j}$ is then calculated by Eq.  \ref{eq:ratio}. 
\end{enumerate}
Thus we generate $B$ (e.g., $B=1000$) random bootstrap samples, and all $B$ bootstrap parameter estimates are ordered from the lowest to highest. Then the   
 $95\%$ percentile bootstrap confidence interval (CI) of the test statistic,  denoted as $(l_{n}, u_{n})$, is obtained such that
\begin{equation}
\vspace{-0.06cm}
   P(  l_{n} < \theta_{i,j}  < u_{n} ) = 95\%
\label{eq:ci}
\vspace{-0.06cm}
\end{equation}
\noindent  e.g., a 95\% percentile bootstrap CI via \numprint{1000} bootstrap samples is the interval between the $25$th quantile and $975$th quantile of \numprint{1000} bootstrap parameter estimates.
If CI $(l_{n}, u_{n})$ does not cover the point $0.0$, we claim the WER difference between two subgroups is statistically significant.
We apply the confidence interval method instead of hypothesis tests – not only for significance tests, but also for quantifying the uncertainty of the test statistic.

\begin{table*}[h!]
\caption{\label{tab:result_fine_tune}{\it  WER results on a test split, after model fine-tuning on a train split. There is no subject of gender ``other'' present in this test subset.}
}
\centerline{ 
\begin{tabular}{  c |  c  |  c c  | c | c c c c  c c  | c }
\hline \hline
      &  overall  &  \multicolumn{3}{c|}{gender} &  \multicolumn{7}{  c}{skin type}    \\ 
model & WER  & female & male  & rel. gap & \Romannum{1} & \Romannum{2} & \Romannum{3} & \Romannum{4} & \Romannum{5} &  \Romannum{6} & rel. gap  \\ 
\hline \hline
Video, supervised  & 11.8 & \textbf{9.8} & \textbf{14.2} & \textbf{45\%}  & \textbf{9.7} & 11.7  & 12.4  & \textbf{12.6} & 11.0 & 11.8 & \textbf{30\%} \\
Video, supervised + fine-tuning & 8.1  & \textbf{6.8} & \textbf{9.6}  & \textbf{42\%} & \textbf{6.7} & 8.1  & \textbf{8.9} & 8.5 & 7.3 & 7.8  & \textbf{33\%} \\
Video, semi-supervised  & 8.4  & \textbf{7.1} & \textbf{9.8}   & \textbf{37\%}   & \textbf{6.9}  & 8.5  & \textbf{9.3}  & 8.8 &  7.4 & 8.0 &  \textbf{35\%} \\ 
Video, semi-supervised + fine-tuning & 7.2 & \textbf{6.0} & \textbf{8.5}   & \textbf{42\%}   & \textbf{6.0} & 7.2 & \textbf{8.0} & 7.6  & 6.4 & 6.9 & \textbf{34\%} \\ 
\hline
\# of hours  &  295   & 161  &  128   &  -     &   11     &   84     &  69  &   25   &  45    & 60    &  --  \\
\hline \hline
\end{tabular}}
\vspace{-0.2cm}
\end{table*}

\section{Results}
\label{sec:results}

The paper that introduced the Casual Conversations data reported on a number of aspects of the metadata: gender, age, skin tone, and lighting conditions \cite{Hazirbas_2021_CVPR}. We report on the same metadata categories except for lighting condition, which we did not expect to impact speech recognition performance.
While skin tone, a purely visual characteristic, is unlikely to directly impact speech recognition performance (as for computer vision models),
we expect that it correlates with other characteristics that may have such an impact, so we opt to report results along that dimension.

\subsection{Evaluating off-the-shelf ASR models}
\label{ssec:results_1}

We first decode the complete 281 hour Casual Conversation audios via each RNN-T model above (Section \ref{sec:model}), and report overall WERs and WERs on each subgroup in Table  \ref{tab:result_all}.  
There appears to be a large performance gap between the female and male speakers with a definite bias towards female speakers, especially for the video models.
We perform the significant tests, as described in Section \ref{sec:bootstrap},  and the WER differences are significant for all models. 
We did not observe much difference across WER by age, except for slightly better performance for the older age category for the LibriSpeech model.

For skin type, we observed noticeable WER differences between various pairs, mostly frequently for the LibriSpeech model.
Then we perform significant tests on the pairwise skin type WERs, and compute the confidence interval (CI) for each test statistic $\theta_{i, j}$ (Eq. \ref{eq:ratio}) of respective subgroup $i$ and $j$ (Section \ref{sec:bootstrap}). 
As shown in Figure \ref{fig:ci}, if the CI does not include point $0.0$, we conclude the WER difference between a skin type pair is significant. Any narrower CI in Figure \ref{fig:ci} indicates a smaller variance in the bootstrap parameter estimates, i.e., WER ratios. The further the complete CI away from point $0.0$, the more significant performance difference is suggested.

We find that the LibriSpeech model has the most occurrences of significant WER differences between subgroups, and video models have the fewest.
We believe that,  for video models, the 2 million hour pseudo-label training data in addition to the transcribed social media videos are more heterogeneous datasets than LibriSpeech, which contains read speech from audiobooks.
Therefore, the training data diversity in social media videos may have helped reduce WER differences between subgroups. 
Note that, although the video teacher model - trained on the same amounts of data as the video student model - provides better overall accuracies, it does not provide more evenly distributed error rates.
Although we do not suggest that skin type has a direct effect on acoustic properties, it may be a proxy for other unobserved characteristics that result in disparate ASR performance.

\subsection{Evaluating the fine-tuned ASR models with in-domain data}
\label{ssec:results_2}

We further use the train/valid/test data split provided in the original dataset release \cite{Hazirbas_2021_CVPR}, to investigate if the unevenly distributed WERs can be reduced by fine-tuning pretrained ASR models with in-domain data\footnote{Given the amount of transcribed speech available, fine-tuning an existing high-performing ASR model provides much better overall accuracies than training a model with Casual Conversations data only. In addition, in line with data use agreement in \cite{Hazirbas_2021_CVPR}, we only use metadata categories (gender, age and skin tone) for ASR model evaluation, and we do not use metadata for any model training purposes.}. 
The training data split is 248 hours in total, and the data size of skin type \Romannum{1} to \Romannum{6} is 10h, 67h, 59h, 22h, 39h, 52h, respectively. The results and test data size are shown in Table \ref{tab:result_fine_tune}.

For both video models of supervised and semi-surprised training, we observe large WER reductions after \numprint{2000} fine-tuning updates. However, in either case, the relative WER differences between subgroups are not reduced, which suggests the model's unbalanced accuracies may not be simply resolved by such in-domain fine-tuning processes.  
Since the relative amounts of data across our metadata categories are unknown for the video dataset, we cannot say that the WER differences between subgroups result from inadequate representation of certain categories of metadata in the broader training data.
However, given (i) the large variability in the million hour video collection, and (ii) the available training data of each skin tone in each train split (used during fine-tuning), we still observe the unfairly distributed errors across metadata.  This suggests there are more fundamental underlying variables associated with the speech styles in the Casual Conversations corpus that deserve additional investigation.

\section{Conclusion}
\label{sec:conclusion}

We have leveraged the existing metadata categories from the Casual Conversations dataset, and then performed an ASR performance evaluation across different models as a function of gender, age, and skin tone categories. Firstly, large accuracy gaps are consistently observed across gender, while no clear bias found towards any age group. 

Secondly, we acknowledge that skin tone is a visual indicator which is suboptimal in measuring an auditory phenomenon.  However, significant WER differences are observed in various comparisons across skin tone categories, suggesting skin tone may correlate with other characteristics that could drive performance differences between subgroups. 
The comparative error rates of various ASR models also indicate that, ASR models trained on sizable and potentially more diverse training data - i.e., the data that more likely contains a diverse range of attributes or subgroup representations - can provide more evenly distributed accuracies, but do not reduce the differences to zero. 

The transcriptions will all be released to the community by the time of paper dissemination; we hope these interesting results inspire the community to continue these investigations to achieve a deeper understanding of the underlying variables affecting speech recognition performance, eventually permitting us to build robust speech systems without having to collect massive amounts of data from each and every target population.

\bibliographystyle{IEEEbib}
\bibliography{strings,refs}

\begin{thebibliography}{10}

\bibitem{buolamwini2018gender}
Joy Buolamwini and Timnit Gebru,
\newblock ``Gender shades: Intersectional accuracy disparities in commercial
  gender classification,''
\newblock in {\em Conference on fairness, accountability and transparency}.
  PMLR, 2018, pp. 77--91.

\bibitem{hbr}
``Voice recognition still has significant race and gender biases,''
  \url{https://hbr.org/2019/05/voice-recognition-still-has-significant-race-and-gender-biases},
\newblock Accessed: 2021-09-10.

\bibitem{forbes}
``Bridging the gender gap in {AI},''
  \url{https://www.forbes.com/sites/falonfatemi/2020/02/17/bridging-the-gender-gap-in-ai/?sh=73c6659e5ee8},
\newblock Accessed: 2021-09-10.

\bibitem{businessinsider}
``{AI} voice recognition racially biased against black voices,''
  \url{https://www.businessinsider.com/study-ai-voice-recognition-racially-biased-against-black-voices-2020-3},
\newblock Accessed: 2021-09-10.

\bibitem{soundhound}
``How to overcome cultural bias in voice {AI} design,''
  \url{https://voices.soundhound.com/how-to-overcome-cultural-bias-in-voice-ai-design},
\newblock Accessed: 2021-09-10.

\bibitem{uxmatters}
``Understanding gender and racial bias in {AI},''
  \url{https://www.uxmatters.com/mt/archives/2020/11/understanding-gender-and-racial-bias-in-ai.php},
\newblock Accessed: 2021-09-10.

\bibitem{racistalgos}
``Racist algorithms,''
  \url{https://medium.com/carre4/racist-algorithms-the-unspoken-bias-in-technology-fa45f309d7c8},
\newblock Accessed: 2021-09-10.

\bibitem{Hazirbas_2021_CVPR}
Caner Hazirbas, Joanna Bitton, Brian Dolhansky, Jacqueline Pan, Albert Gordo,
  and Cristian~Canton Ferrer,
\newblock ``Casual conversations: A dataset for measuring fairness in {AI},''
\newblock in {\em Proceedings of the IEEE/CVF Conference on Computer Vision and
  Pattern Recognition (CVPR) Workshops}, June 2021, pp. 2289--2293.

\bibitem{dwork2012fairness}
Cynthia Dwork, Moritz Hardt, Toniann Pitassi, Omer Reingold, and Richard Zemel,
\newblock ``Fairness through awareness,''
\newblock in {\em Proceedings of the 3rd Innovations in Theoretical Computer
  Science Conference}, 2012, pp. 214--226.

\bibitem{chouldechova2017fair}
Alexandra Chouldechova,
\newblock ``Fair prediction with disparate impact: A study of bias in
  recidivism prediction instruments,''
\newblock {\em Big data}, vol. 5, no. 2, pp. 153--163, 2017.

\bibitem{chouldechova2017fairer}
Alexandra Chouldechova and Max G'Sell,
\newblock ``Fairer and more accurate, but for whom?,''
\newblock {\em arXiv preprint arXiv:1707.00046}, 2017.

\bibitem{tatman2017gender}
Rachael Tatman,
\newblock ``Gender and dialect bias in {YouTube’s} automatic captions,''
\newblock in {\em Proceedings of the First ACL Workshop on Ethics in Natural
  Language Processing}, 2017, pp. 53--59.

\bibitem{koenecke2020racial}
Allison Koenecke, Andrew Nam, Emily Lake, Joe Nudell, Minnie Quartey, Zion
  Mengesha, Connor Toups, John~R Rickford, Dan Jurafsky, and Sharad Goel,
\newblock ``Racial disparities in automated speech recognition,''
\newblock {\em Proceedings of the National Academy of Sciences}, vol. 117, no.
  14, pp. 7684--7689, 2020.

\bibitem{garnerin2019gender}
Mahault Garnerin, Solange Rossato, and Laurent Besacier,
\newblock ``{Gender representation in French Broadcast Corpora and its impact
  on ASR performance},''
\newblock in {\em Proceedings of the 1st International Workshop on AI for Smart
  TV Content Production, Access and Delivery}, 2019, pp. 3--9.

\bibitem{feng2021quantifying}
Siyuan Feng, Olya Kudina, Bence~Mark Halpern, and Odette Scharenborg,
\newblock ``Quantifying bias in automatic speech recognition,''
\newblock {\em arXiv preprint arXiv:2103.15122}, 2021.

\bibitem{garnerin2021investigating}
Mahault Garnerin, Solange Rossato, and Laurent Besacier,
\newblock ``Investigating the impact of gender representation in asr training
  data: a case study on librispeech,''
\newblock {\em GeBNLP 2021}, p.~86, 2021.

\bibitem{panayotov2015librispeech}
Vassil Panayotov, Guoguo Chen, Daniel Povey, and Sanjeev Khudanpur,
\newblock ``{LibriSpeech}: an {ASR} corpus based on public domain audio
  books,''
\newblock in {\em Proc. ICASSP}, 2015.

\bibitem{sari2021counterfactually}
Leda Sari, Mark Hasegawa-Jonson, and Chang-D Yoo,
\newblock ``Counterfactually fair automatic speech recognition,''
\newblock {\em IEEE Transactions on Audio, Speech \& Language Processing}, 2021
  (accepted).

\bibitem{kusner2017counterfactual}
Matt~J Kusner, Joshua~R Loftus, Chris Russell, and Ricardo Silva,
\newblock ``Counterfactual fairness,''
\newblock {\em arXiv preprint arXiv:1703.06856}, 2017.

\bibitem{kendall2020corpus}
Tyler Kendall and Charlie Farrington,
\newblock ``The corpus of regional {African American Language},''
\newblock {\em The Online Resources for African American Language Project},
  vol. Version 2020.05, 2020.

\bibitem{meyer2020artie}
Josh Meyer, Lindy Rauchenstein, Joshua~D Eisenberg, and Nicholas Howell,
\newblock ``Artie bias corpus: An open dataset for detecting demographic bias
  in speech applications,''
\newblock in {\em Proceedings of The 12th Language Resources and Evaluation
  Conference}, 2020, pp. 6462--6468.

\bibitem{graves2012sequence}
Alex Graves,
\newblock ``Sequence transduction with recurrent neural networks,''
\newblock {\em arXiv preprint arXiv:1211.3711}, 2012.

\bibitem{gulati2020conformer}
Anmol Gulati, James Qin, Chung-Cheng Chiu, Niki Parmar, Yu~Zhang, Jiahui Yu,
  Wei Han, Shibo Wang, Zhengdong Zhang, Yonghui Wu, et~al.,
\newblock ``Conformer: Convolution-augmented transformer for speech
  recognition,''
\newblock in {\em Proc. Interspeech}, 2020.

\bibitem{yeh2021streaming}
Ching-Feng Yeh, Yongqiang Wang, Yangyang Shi, Chunyang Wu, Frank Zhang, Julian
  Chan, and Michael~L Seltzer,
\newblock ``Streaming attention-based models with augmented memory for
  end-to-end speech recognition,''
\newblock in {\em Proc. SLT}, 2021.

\bibitem{kudo2018sentencepiece}
Taku Kudo and John Richardson,
\newblock ``Sentencepiece: A simple and language independent subword tokenizer
  and detokenizer for neural text processing,''
\newblock {\em arXiv preprint arXiv:1808.06226}, 2018.

\bibitem{li2021better}
Bo~Li, Anmol Gulati, Jiahui Yu, Tara~N Sainath, Chung-Cheng Chiu, Arun
  Narayanan, Shuo-Yiin Chang, Ruoming Pang, Yanzhang He, James Qin, et~al.,
\newblock ``A better and faster end-to-end model for streaming {ASR},''
\newblock in {\em Proc. ICASSP}, 2021.

\bibitem{shi2021emformer}
Yangyang Shi, Yongqiang Wang, Chunyang Wu, Ching-Feng Yeh, Julian Chan, Frank
  Zhang, Duc Le, and Mike Seltzer,
\newblock ``Emformer: Efficient memory transformer based acoustic model for low
  latency streaming speech recognition,''
\newblock in {\em Proc. ICASSP}, 2021.

\bibitem{park2019specaugment}
Daniel~S Park, William Chan, Yu~Zhang, Chung-Cheng Chiu, Barret Zoph, Ekin~D
  Cubuk, and Quoc~V Le,
\newblock ``{SpecAugment}: A simple data augmentation method for automatic
  speech recognition,''
\newblock {\em Proc. Interspeech}, 2019.

\bibitem{mahadeokar2021alignment}
Jay Mahadeokar, Yuan Shangguan, Duc Le, Gil Keren, Hang Su, Thong Le,
  Ching-Feng Yeh, Christian Fuegen, and Michael~L Seltzer,
\newblock ``Alignment restricted streaming recurrent neural network
  transducer,''
\newblock in {\em Proc. SLT}, 2021.

\bibitem{liu2021improving}
Chunxi Liu, Frank Zhang, Duc Le, Suyoun Kim, Yatharth Saraf, and Geoffrey
  Zweig,
\newblock ``Improving rnn transducer based asr with auxiliary tasks,''
\newblock in {\em Proc. SLT}, 2021.

\bibitem{ott2019fairseq}
Myle Ott, Sergey Edunov, Alexei Baevski, Angela Fan, Sam Gross, Nathan Ng,
  David Grangier, and Michael Auli,
\newblock ``fairseq: A fast, extensible toolkit for sequence modeling,''
\newblock in {\em Proceedings of NAACL-HLT 2019: Demonstrations}, 2019.

\bibitem{kingma2014adam}
Diederik~P Kingma and Jimmy Ba,
\newblock ``Adam: A method for stochastic optimization,''
\newblock in {\em Proc. ICLR}, 2015.

\bibitem{115546}
D.S. Pallet, W.M. Fisher, and J.G. Fiscus,
\newblock ``Tools for the analysis of benchmark speech recognition tests,''
\newblock in {\em Proc. ICASSP}, 1990.

\bibitem{266481}
L.~Gillick and S.J. Cox,
\newblock ``Some statistical issues in the comparison of speech recognition
  algorithms,''
\newblock in {\em Proc. ICASSP}, 1989.

\bibitem{efron1994introduction}
Bradley Efron and Robert~J Tibshirani,
\newblock {\em An introduction to the bootstrap},
\newblock CRC press, 1994.

\end{thebibliography}

\end{document}